\documentclass[letterpaper,twocolumn,10pt]{article}
\usepackage{usenix2022_SOUPS}
\usepackage{dirtytalk}
\usepackage{graphicx}
\usepackage{url}

\usepackage{breakurl}

\usepackage{tikz}
\usepackage{amsmath}

\usepackage{filecontents}

\begin{document}
%-------------------------------------------------------------------------------

%don't want date printed
\date{}

% make title bold and 14 pt font (Latex default is non-bold, 16 pt)
\title{\Large \bf Inclusive Privacy Design for Older Adults Living in Ambient Assisted Living}

% if you leave this blank it will default to a possibly ugly attempt 
% to make the contents of the \author command below into a string
\def\plainauthor{Author name(s) for PDF metadata. Don't forget to anonymize for submission!}

%for single author (just remove % characters)
\author{
% {\rm Nyteisha\ Bookert} \\
% North Carolina Agricultural \& Technical State University
% \and
% {\rm May\ Almousa}\\
% North Carolina Agricultural \& Technical State University
% \and
% {\rm Mohd Anwar}\\
% \textcolor{red}{North Carolina Agricultural \& Technical State University}
% \and
{\rm Nyteisha\ Bookert,\ May\ Almousa,\ and\ Mohd Anwar}\\
{Department of Computer Science , College of Engineering}\\
{North Carolina A\&T State University}\\ 
{North Carolina, USA}
%{Computer Science Department}\\
%{North Carolina Agricultural \& Technical State University}\\ 
%{Greensboro, NC, USA}
%\orgdiv{Computer Science Department, College of Engineering}, \orgname{North Carolina A\&T State University}, \orgaddress{\state{North Carolina}, \country{USA}}
% copy the following lines to add more authors
% \and
% {\rm Name}\\
%Name Institution
} % end author

\maketitle
\thecopyright

%-------------------------------------------------------------------------------
\begin{abstract}
%-------------------------------------------------------------------------------
Ambient assisted living (AAL) environments support independence and quality of life of older adults However, in an AAL environment, privacy-related issues (e.g., unawareness, information disclosure, and lack of support) directly impact older adults and bystanders (e.g., caregivers, service providers, etc.). We explore the privacy challenges that both older adults and bystanders face in AAL. We call for inclusive privacy design and recommend following areas of improvement: consent, notification, and consideration for cultural differences. 
\end{abstract}

%-------------------------------------------------------------------------------
\section{Introduction}
The aging population of the world is rapidly growing due to an increased life expectancy and declining fertility rates. Individuals 65 and older are expected to outnumber children under five for the first time in history \cite{NIHreport}. This demographic shift raises several issues, such as an increase in age-related diseases and health care costs, a shortage of caregivers, and a rise in dependency \cite{Rashidi}. We need solutions that reduce the burden on society and enable older adults to manage their health and safely maintain their autonomy. 

The Internet of Medical Things (IoMT) enables medical adherence, activity recognition, and disease management \cite{FARAHANI2018659, Habibzadeh}. Furthermore, IoMT supports remote health monitoring to share health-related and other relevant information with caregivers, healthcare providers, and family members. IoMT reduces the strain of nursing homes by supporting Ambient Assisted Living, allowing elderly adults to remain in their own homes \cite{8250384}, \cite{Dohr}. AAL environments contain different IoMT devices and other technologies that use interrelated computing devices to transfer data without requiring human-to-computer interaction to facilitate a seamless living environment. These technologies include emergency help systems, fall detection systems, and vital signs monitoring.

When used correctly, the emerging technology supports older adults to be more organized \cite{peek2016older}. Everyday tasks, such as cleaning, can become challenging for seniors, technologies including robot vacuums, can make tasks more manageable. Other technologies such as automated lights can significantly lessen the odds of injury and risk of falls at night. The adoption of innovative home technology can support aging at home. Intelligent devices offer an alternative, less costly solution by leveraging technological developments. In addition, while preserving their autonomy, older adults can deepen their relationships with friends and family through smart home technology \cite{heinz2013perceptions}.

However, IoMT and AAL environments expose user data to the risk of known networking vulnerabilities and introduce new threats against patients, devices, and medical infrastructure. Hence, data privacy stands as one of the barriers to adoption \cite{8923448}. Privacy concerns and issues must be addressed in the design process. Moreover, privacy preferences and expectations are diverse. We explore the privacy challenges that older adults and bystanders face in AAL. To address these challenges, we propose a research agenda with the following research questions. What privacy issues directly impact older adults and bystanders in an AAL environment? How do cultural differences impact the IoMT settings in a smart home (or AAL environment)? Then we suggest further evaluating how effective are current privacy-enhancing technologies to addressing the issues identified through the first two questions.

The remainder of this paper is organized as follows. We introduce the older adult population group and their needs for IoMT in section 2. Section 3 presents the AAL scenario. A discussion on the privacy challenges is in section 4, and key considerations for inclusive design are in section 5. Finally, we summarize our recommendations in section 6.

%-------------------------------------------------------------------------------
\section{Adoption of IoMT among Older Adult Population Group}
%-------------------------------------------------------------------------------
According to the American Association of Retired Persons (AARP)~\cite{AARP}, in the US, about 90 percent of adults over 65 prefer to age at home. This creates an enormous demand for IoMT devices. The IoMT devices with the software and health systems form an ecosystem that improves patient care and helps streamline the workflow in the healthcare systems. Patients are, through these technologies, able to save on medical-related costs and have a timely intervention on their health conditions~\cite{razdan2021internet}. According to Khan et al. (2021)~\cite{khan2021iomt}, coupling healthcare with IoMT facilitates care administration and the quality of life while still making the frameworks cost-effective.

The ability to track key health indicators remotely became especially critical during the peak of the COVID-19 pandemic. Now, older adults and their caregivers increasingly rely on remote health monitoring to stay in their homes safely.

As the “Care Anywhere” healthcare trend continues with IoMT technology, elderly people are left particularly vulnerable to privacy issues. Because most older adults are overburdened with their health concerns, they tend to use any IoMT device recommended to them. In many cases, the devices take their privacy and control away, which is especially concerning in a population that is less technologically literate. Yet, IoMT helps the aging population remain independent for as long as possible.

%-------------------------------------------------------------------------------
\section{Ambient Assisted Living Environment}
\label{sec:figs}

%-------------------------------------------------------------------------------

Ambient Assisted Living (AAL) uses technologies to enable and support older adults and those with special needs to live independently and provide more autonomy. Health and activity monitoring sensors, such as biosensors (heart rate, blood pressure, body temperature) and environmental (temperature, light, humidity), allow for continuous health and cognitive status evaluation. Evaluating the changes in activities or abnormal activities may suggest a health decline. PIR, RFID, motion sensors, and GPS helps track and monitor older adults’ movements throughout the space and enables wandering prevention tools. For example, if an individual approaches a potentially dangerous area, the relevant parties can receive an alert to intervene. A caregiver can monitor audio and video with microphones and surveillance cameras, verify medication adherence, and offer automated planning and scheduling \cite{Lloret,Rashidi}. The environment includes multiple wireless access points and a smartphone with an AAL App to manage the various sensors, actuators, and devices. AAL technology can be implemented in traditional residential homes, nursing homes, or outpatient facilities. Next, we provide three use cases in the AAL environment. 

\begin{figure}
\begin{center}
\includegraphics[width=8.5cm,scale=1]{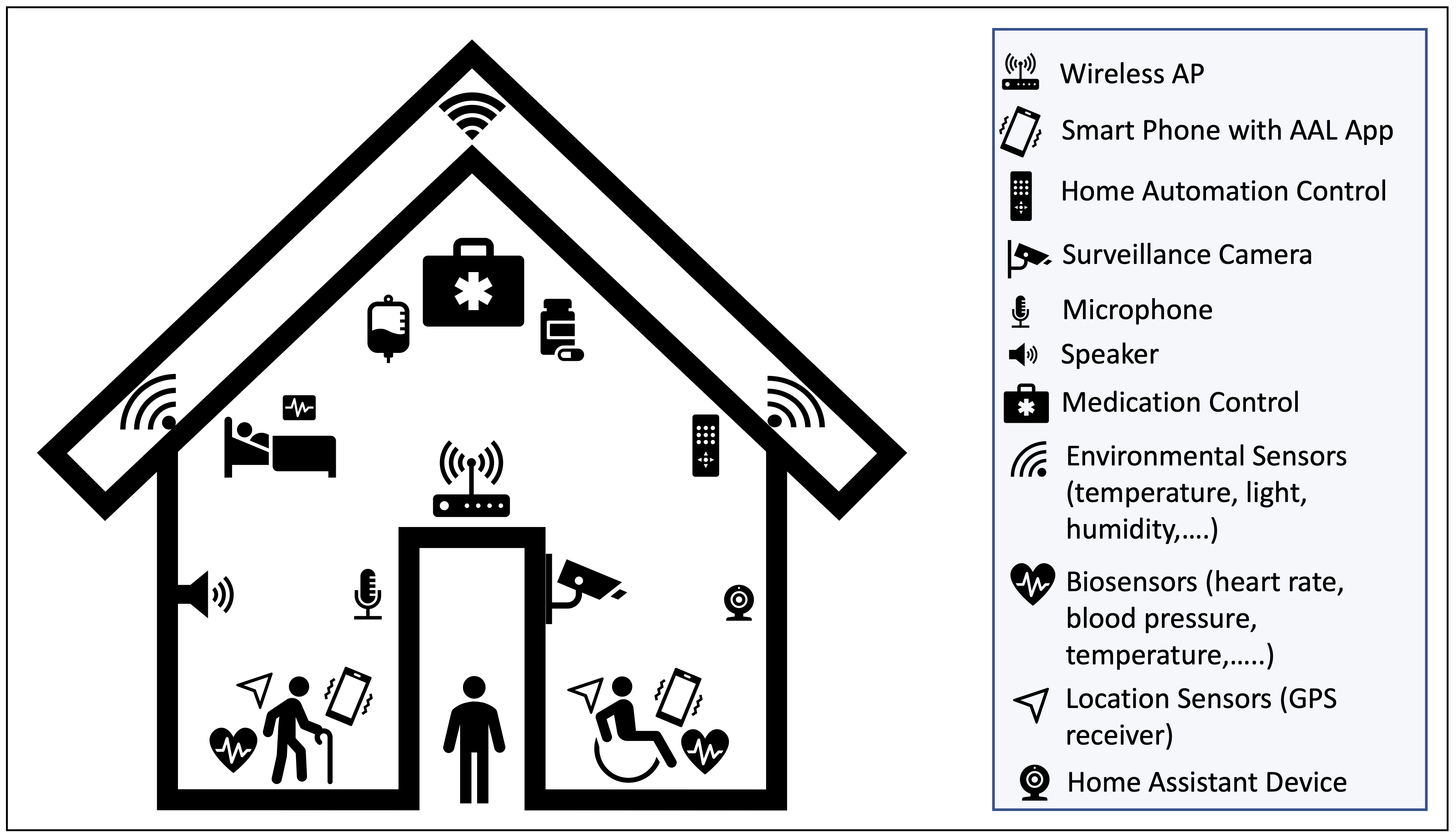}

\caption{\label{AAL_house} Ambient Assisted Living Environment}
%\cite{Lloret}
\end{center}
\end{figure}

\begin{enumerate}
    % \item The use case should show the privacy implications of AAL technologies. Currently looking for a use case that will highlight the three issues listed below. 
    \item An adult child procures an AAL home for their older parent, as shown in Figure\ref{AAL_house}. The child explains to the parent the capabilities of the home, such as being able to control home automation (e.g., opening and closing the blinds, the lights automatically turning on when entering a room), track medication usage, and detect emergencies (e.g., detects falls, detects no movement when movement is expected). The child helps the parent access the app but does not detail the types of devices that support and enable those applications in the home.
    \item After moving into the home, the older adult has a caregiver who visits the home daily to assist with grocery shopping, errands, laundry, and light housekeeping tasks.
    \item Each week, the older adult hosts a game night with a few friends who bring a dish to share. They use the kitchen area to store the food and occasionally use the restroom.
\end{enumerate}

%-------------------------------------------------------------------------------
\section{Privacy Concerns/Challenges}
%-------------------------------------------------------------------------------
\subsection{Bystanders}
The literature has established that smart technologies have primarily focused on end-users without considering other individuals, such as bystanders affected by the privacy settings and postures of the IoMT devices. The bystanders are not either target users or owners of the devices. Yao et al. \cite{yao2019privacy} define bystanders as individuals \say{who do not own or directly use these devices but are potentially involved in the use of smart home devices, such as other family members who do not purchase the devices, guests, tenants, passersby, etc.} Besides, in an AAL environment, an older adult or a person with special needs may not have decided to purchase the IoMT device or own the living environment, which creates yet another unique scenario for bystanders to consider. Even though older adults directly interact with and use these devices, they may not have insight or control over the use scenarios or privacy configurations. Consideration must be given to the interaction patterns between target end-users and bystanders -- potential guests or visitors (e.g., caregivers, visiting friends, or family). 

\subsection{Unawareness}
In an AAL system, various sensors, actuators, and devices collect health information. Traditionally, these devices operate in the background without human interaction, creating the risk of unawareness. Unawareness refers to a person not knowing about the collection and sharing practices or purposes of a system \cite{deng2011privacy}. In an AAL environment, the risk is compounded for bystanders and future users. It is imperative to consider how an individual gains access to an AAL environment and the impact it may have on their privacy. For example, someone who moves into an AAL home may not know or remember each item (sensors, actuators, devices). It may be difficult for an occupant to reveal the devices, sensors, and actuators making the technology possible. For instance, an occupant may understand that a system monitors their recovery after knee surgery but cannot discover which elements support the process. Additional challenges include discovering the privacy settings for the system and locating privacy notices of the respective device - which may fail to support users’ preferences. Cultural differences may further expand the misconceptions about the data flow. Individuals from a collectivism \cite{hofstede2001culture} culture may expect their family members to have the ability to monitor their day-to-day activities, while others may find this intrusive.

\subsection{Information Disclosure}
Disclosure of information is the ability of an unauthorized party or system to learn the contents or information about a person or entity ~\cite{Ziegeldorf_2013}. Some examples of the disclosure include a nurse accessing records of a patient not under their charge, an adversary gaining access to pharmaceutical research data, and a company selling data without the consumers' permission. Additionally, IoT devices may contain privacy-violating interactions and presentations that reveal information to others. The definition of an authorized person may be unclear in an AAL environment. Consider the adult child and older adult in scenario 1. Who decides who has access to the older adults' information? Is it the person who creates the account for the older adult or the older adult who will live in the environment? If the adult child is not an authorized user but gains access to the older adults' system and eavesdrops during the weekly game night, it is considered unauthorized disclosure.

\subsection{Need for Support}
The use of IoMT devices poses some challenges for older adults because they have limited technical experiences, knowledge, and declining physical and cognitive abilities. These are attributes that put older adults at a higher risk with the use of this technology. The ever-changing nature of the cybersecurity and technology landscape, which are designed with less consideration of their age group, leads to some level of mismatched mental models causing challenges such as privacy and usability\cite{pakianathan2020towards}. Another issue is that older adult tends to rely on other people for many things due to their age-related challenges, including the operation or interaction with technological devices such as the IoMT\cite{pirhonen2020these}. Therefore, while seeking help on the devices in the AAL environment, their health information which should be private can be exposed to extended family members, neighbors, or caregivers.   

The limited knowledge regarding technical issues such as privacy settings of an IoMT device makes the elderly less concerned about these aspects compared to other groups. As a result, even when their privacy got compromised while using IoMT devices, they will not be able to address the problem. Yao et al. (2019)\cite{yao2019privacy} argue that older adults do not engage in privacy-seeking behaviors to mitigate privacy concerns. They tend to place a high degree of trust in the manufacturers of devices and systems, which leaves them with the perception that the developers of devices will factor in all their privacy and safety needs.

The IoMT devices also have a design challenge that makes it hard to maintain privacy. IoMT devices pose serious risks in terms of privacy for older adults who are less aware, inexperienced, and less concerned with the risks involved. There should be considerations from the device developers in dealing with privacy issues. The privacy control settings of these devices are left to the users by default. Whereas, in most cases, the older adults age group shares these devices with other users, especially those supporting them. Therefore, the design of IoMT needs to consider the socio-cultural factors in the privacy design for the older adults age group.

\section{Call for Inclusive Privacy Design in AAL/IoMT}

%-------------------------------------------------------------------------------
\begin{enumerate}
    \item \textbf{\textit{Consent + Data Collection.}} As a result of using the IoMT devices, a trove of data is collected by the device manufacturers. The collected data is used to assess the health status of the user and offer a suitable and immediate response by the device. Currently, significant research around privacy concerns has focused on end-users of smart home devices. For many older adults, privacy concerns are overshadowed by a strong desire to remain independent while keeping their health under control. For IoMT, the industry must go beyond just users’ privacy expectations. Desired privacy mechanisms must also cover the bystander’s perspective. Even though the bystanders are not the owner or primary users of the device, they may be affected through data collection settings. Because house guests and other family members have not consented to data collection, privacy issues may arise. Even when an older adult may not have privacy concerns, IoMT devices must consider the possibility of accidental data collection from bystanders or conflicting privacy preferences of a bystander.
    
    Designers should consider implementing cooperative mechanisms between owners and bystanders, such as easy access for users and bystanders, and provide bystanders an opportunity to share privacy preferences with owners. There is an increased need for privacy awareness at the smart home level compared to the device level \cite{10.1145/3491102.3502137}. Furthermore, bystander-centric mechanisms are necessary to aid bystanders in detecting nearby devices, revealing device behaviors, limiting data collection, and controlling personal data \cite{yao2019privacy}. The IoT Privacy Infrastructure \cite{8490188} serves as an example at a large scale. 
    % Add the computer vision information to this area.
    
    \item \textbf{\textit{Notifications.}} Notifications are typical features of IoMT devices and AAL systems. For example, a user may receive alerts to take their medication or to exercise. If a system detects a fall, the user may receive a prompt to determine whether the incident was a false alarm or an actual event. Emergency contacts and emergency personnel may receive an alert during a confirmed incident. There are several ways to present notifications within the environment. However, they introduce unique potential privacy concerns. For example, the medication reminder could be a visual notification on the smartphone or an audio reminder shared in the room the person is currently occupying. If the audio reminder shares the information while a guest is present, it may reveal sensitive information. An appropriate level of specificity should be included in the notification that provides appropriate details to the individual but does not disclose sensitive information to bystanders. Mechanisms should also be in place to allow the user to silence notifications to limit potential exposure. Additionally, devices and sensors in an AAL environment that collect sensitive data (audio, video, and health information) should notify users of their presence with indicators. Indicators range from blinking lights on a camera to an audible before a recording begins. However, the indicator's design should be inclusive to accommodate individual needs and consider unobtrusive modalities \cite{10.1145/3491102.3502137}.

    \item \textbf{\textit{Cultural Differences.}} System designers should consider how demographic and other cultural factors (e.g., age, relationship dynamics, sexuality, norms, race, gender) alter preferences and expectations of the environment. For example, the parent-child relationship scenario introduces several privacy threats, such as unawareness about specific devices and sensors in the environment and thereby disclosure of information. Yet, the dynamic of that relationship and cultural norms may impact how the parent views this interaction. The parent may want to know more about the environment. Flexible controls should be in place for flexibility depending on the relationship dynamics and accepted cultural beliefs. Furthermore, the proper controls may also benefit other bystanders as they navigate the home. Parents may also choose to limit access to their information. The controls should enable bystanders and users to balance privacy and other key factors, such as safety and health.  

\end{enumerate}

%-------------------------------------------------------------------------------

\section{Summary}
%-------------------------------------------------------------------------------
As IoMT and AAL emerge as solutions to mitigate the challenges of the aging population, it is imperative to identify and address the privacy concerns of all individuals, including bystanders. By considering sample scenarios of bystanders in AAL environments interacting with IoMT devices, we establish the need for the research community to investigate the privacy issues directly impacting bystanders in an AAL environment and how cultural differences affect IoMT settings in an AAL environment. Yet, effective solutions must account for challenges such as unawareness and reliance on others for support and consider cultural differences to provide better tools. Furthermore, we must examine how effectively current privacy-enhancing technologies reduce privacy risks in AAL environments.

%-------------------------------------------------------------------------------
\section*{Acknowledgments}
%-------------------------------------------------------------------------------
This work was supported partially by the U. S. Department of Education under the Title III Historically Black Graduate Institutions (HBGI) grant. The views and conclusions contained herein are those of the authors and should not be interpreted as necessarily representing the official policies or endorsements, either expressed or implied, of the U. S. Government. The U. S. Government is authorized to reproduce and distribute reprints for governmental purposes notwithstanding any copyright annotation therein.

----------

%Here's a typical reference to a floating figure: Figure~\ref{fig:vectors}. Floats should usually be placed where latex wants then. Figure\ref{fig:vectors} is centered, and has a caption that instructs you to make sure that the size of the text within the figures that you use is as big as (or bigger than) the size of the text in the caption of the figures. Please do. Really.

%In our case, we've explicitly drawn the figure inlined in latex, to allow this tex file to cleanly compile. But usually, your figures will reside in some file.pdf, and you'd include them in your document with, say, \textbackslash{}includegraphics.

%Lists are sometimes quite handy. If you want to itemize things, feel free:

%\begin{description}
  
%\item[fread] a function that reads from a \texttt{stream} into the
 % array \texttt{ptr} at most \texttt{nobj} objects of size
 % \texttt{size}, returning returns the number of objects read.

%\item[Fred] a person's name, e.g., there once was a dude named Fred
 % who separated usenix.sty from this file to allow for easy
%  inclusion.
%\end{description}

%\noindent
%The noindent at the start of this paragraph in its tex version makes
%it clear that it's a continuation of the preceding paragraph, as
%opposed to a new paragraph in its own right.

%\subsection{LaTeX-ing Your TeX File}
%-----------------------------------

%-------------------------------------------------------------------------------

%-------------------------------------------------------------------------------
\bibliographystyle{plain}
\bibliography{usenix2022_SOUPS.bib}

%%%%%%%%%%%%%%%%%%%%%%%%%%%\jobname%%%%%%%%%%%%%%%%%%%%%%%%%%%%%%%%%%%%%%%%%%%%%%%%%%%%%
\end{document}